\documentclass[prl,twocolumn,showpacs,floats]{revtex4}

\usepackage{graphicx}
\begin{document}

%
\title{Extraction of Electron Self-Energy and Gap Function in the Superconducting State of
Bi$_2$Sr$_2$CaCu$_2$O$_{8}$ Superconductor via Laser-Based Angle-Resolved Photoemission}
%
%
%
\author{Wentao Zhang$^{1}$, Jin Mo Bok$^{2}$, Jae Hyun Yun$^{2}$, Junfeng He$^{1}$, Guodong Liu$^{1}$, Lin Zhao$^{1}$, Haiyun Liu$^{1}$, Jianqiao Meng$^{1}$, Xiaowen Jia$^{1}$, Yingying Peng$^{1}$, Daixiang Mou$^{1}$, Shanyu Liu$^{1}$, Li Yu$^{1}$,  Shaolong He$^{1}$, Xiaoli Dong$^{1}$, Jun Zhang$^{1}$, J. S. Wen$^{3}$, Z. J. Xu$^{3}$, G. D. Gu$^{3}$,  Guiling Wang$^{4}$, Yong Zhu$^{4}$, Xiaoyang Wang$^{4}$, Qinjun Peng$^{4}$, Zhimin Wang$^{4}$, Shenjin Zhang$^{4}$, Feng Yang$^{4}$, Chuangtian Chen$^{4}$, Zuyan Xu$^{4}$, H.-Y. Choi$^{2}$, C. M. Varma$^{5}$ and X. J. Zhou $^{1,*}$}

\affiliation{
\\$^{1}$National Laboratory for Superconductivity, Beijing National Laboratory for Condensed Matter Physics, Institute of Physics, Chinese Academy of Sciences, Beijing 100080, China
\\$^{2}$Department of Physics and Institute for Basic Science Research, SungKyunKwan University, Suwon 440-746, Korea.
\\$^{3}$Condensed Matter Physics and Materials Science Department, Brookhaven National Laboratory, Upton, New York 11973, USA
\\$^{4}$Technical Institute of Physics and Chemistry, Chinese Academy of Sciences, Beijing 100080, China
\\$^{5}$Department of Physics and Astronomy, University of California, Riverside, California 92521
}
\date{March 18, 2011}

\begin{abstract}
Super-high resolution laser-based angle-resolved photoemission measurements have been performed on a high temperature superconductor Bi$_2$Sr$_2$CaCu$_2$O$_8$. The band back-bending characteristic of the Bogoliubov-like quasiparticle dispersion is clearly revealed at low temperature in the superconducting state. This makes it possible for the first time to experimentally extract the complex electron self-energy and the complex gap function in the superconducting state.  The resultant electron self-energy and gap function exhibit features at $\sim$54 meV and $\sim$40 meV, in addition to the superconducting gap-induced structure at lower binding energy and a broad featureless structure at higher binding energy.  These information will provide key insight and constraints on the origin of electron pairing in high temperature superconductors.
\end{abstract}

\pacs{74.72.Gh, 74.25.Jb, 79.60.-i, 74.20.Mn}

\maketitle

The mechanism of high temperature superconductivity in the copper-oxide compounds (cuprates) remains an outstanding issue in condensed matter physics after its first discovery more than two decades ago\cite{Bednorz}.  It has been established that, in cuprate superconductors, the electrons are paired with opposite spins and opposite momentum to form a spin-singlet state\cite{CEGough}, as the Cooper pairing in the conventional superconductors\cite{BCSTheory}. It has been further established that the superconducting gap in the cuprate superconductors has a predominantly {\it d}-wave symmetry\cite{TunnelingJunction}, distinct from the {\it s}-wave form in the conventional superconductors.  The center of debate lies in the origin of the electron pairing in the cuprate superconductors. In the conventional superconductors, it is known from the BCS  theory of superconductivity that the exchange of phonons  gives rise to the formation of Cooper pairing\cite{BCSTheory}. In the cuprate superconductors, the question becomes whether the electron-electron interaction can automatically cause the pairing\cite{AndersonScience}, or a distinct collective mode (a glue) remains essential in mediating the pairing as in the conventional superconductors and what the nature of the glue is\cite{Scalapino, ZXShen, Aji}.

In the conventional superconductors, the extraction of the electron self-energy  $\Sigma(\omega)$ and gap function  $\phi(\omega)$ in the superconducting state played a critical role in proving that phonons provide the glue for Cooper pairs\cite{ScalapinoPark}. Experimentally, these two fundamental quantities were extracted from the tunneling experiments\cite{IGiaever}. On the one hand, theoretical models were examined by simulating the two quantities to compare with the experimentally extracted ones\cite{SchriefferSimu}. On the other hand, the underlying bosonic spectral function associated with the pairing glue was directly inverted from the gap function via the Eliashberg equations\cite{McMillanInversion}.  The striking resemblance between the bosonic spectral function thus extracted and the phonon density of states directly measured  from the neutron scattering provided an overwhelming evidence of the phonons as a pairing glue in the conventional superconductors\cite{ScalapinoPark}. It is natural to ask whether similar procedures can be applied in high temperature cuprate superconductors, which necessitates reliable extraction of the electron self-energy and gap function in the superconducting state\cite{Vekhter}. However, the ordinary tunneling experiments cannot be used for this purpose due to the complications from the strong anisotropy of the electronic structure and {\it d}-wave superconducting gap\cite{HTSCTunneling,DavisSTM}.

\begin{figure}[tbp]
\begin{center}
\includegraphics[width=1.00\columnwidth,angle=0]{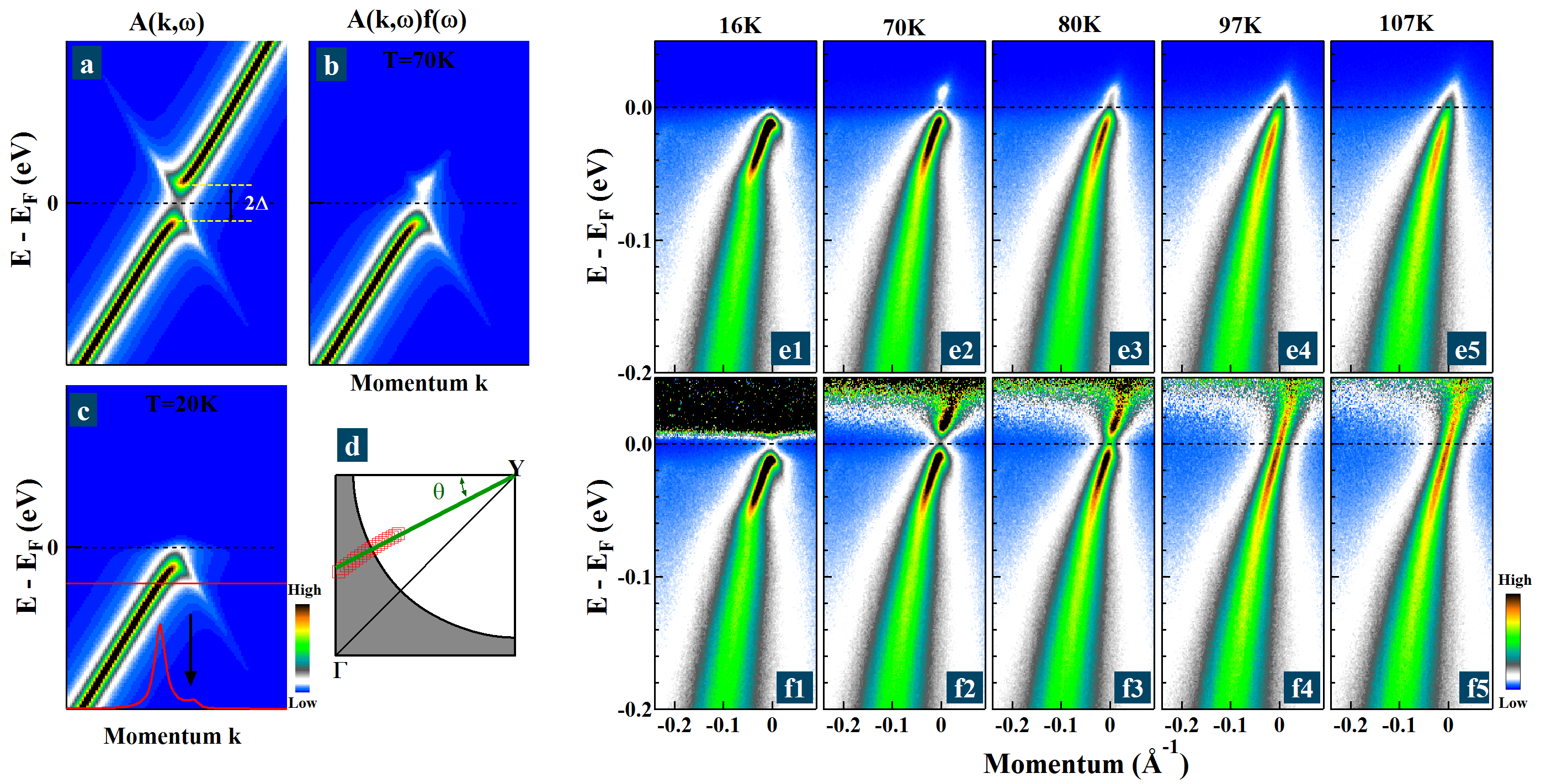}
\end{center}
\caption{Observation of two branches of Bogoliubov quasiparticle-like dispersions in the superconducting state of the slightly underdoped Bi2212 (T$_c$=89 K). (a). Simulated single particle spectral function A(k,$\omega$) in the superconducting state, taking a superconducting gap of $\Delta$=10 meV and a linewidth broadening of $\Gamma$=5 meV. The simulated A(k,$\omega$) multiplied by the Fermi distribution function f($\omega$) at a temperature of 70 K is shown in (b) and at 20 K is shown in (c).  (e1-e5). Photoemission images taken at different temperatures along the momentum cut marked as in (d).  This cut nearly points towards ($\pi$, $\pi$) with an angle $\theta$ as defined.  (f1-f5). Photoemission images in (e1-e5) divided by a Fermi distribution function at the corresponding temperatures. }
\end{figure}

In this paper, we report the first experimental extraction of  the electron self-energy $\Sigma(\omega)$ and gap function $\phi(\omega)$ in the superconducting state of a high temperature superconductor Bi$_2$Sr$_2$CaCu$_2$O$_{8}$ (Bi2212) from angle-resolved photoemission (ARPES) measurements.  This is accomplished by carrying out super-high resolution laser-based ARPES measurements on Bi2212 which clearly reveals the band back-bending behavior of the Bogoliubov-like quasiparticle dispersion at low temperature. The extracted $\Sigma(\omega)$ and $\phi(\omega)$ show clear features at $\sim$54 meV and $\sim$40 meV,  in addition to the superconducting gap-induced structure at lower energies, and a broad background at higher energies.  The accurate and detailed determination of $\Sigma(\omega)$ and  $\phi(\omega)$ will  provide key information on the pairing mechanism of high temperature superconductivity.

The angle-resolved photoemission measurements were carried out on our vacuum ultra-violet (VUV) laser-based ARPES system\cite{LiuIOP}. The photon energy of the laser is 6.994 eV with a bandwidth of 0.26 meV. The energy resolution of the electron energy analyzer (Scienta R4000) was set at 1 meV, giving rise to an overall energy resolution of $\sim$1.2 meV which is significantly improved from 7$\sim$15 meV  from some previous ARPES measurements\cite{HMatsui_Bi2223,AVB_BogoAngle,WSLee,JC_Bendingback,HBYang_CooperPair}. The angular resolution is $\sim$0.3$^\circ$, corresponding to a momentum resolution  $\sim$0.004 $\AA$$^{-1}$ at the photon energy of 6.994 eV, more than twice improved from 0.009 $\AA$$^{-1}$ at a regular photon energy of 21.2 eV for the same angular resolution. The Fermi level is referenced by measuring on a clean polycrystalline gold that is electrically connected to the sample. Slightly underdoped Bi2212 (T$_c$=89 K) single crystals were cleaved {\it in situ} and measured in vacuum with a base pressure better than 5 $\times$ 10$^{-11}$ Torr.

According to the BCS theory of superconductivity\cite{BCSTheory}, in the superconducting state, the low energy excitations are Bogoliubov quasiparticles representing the coherent mixture of electron and hole components. The BCS spectral function  can be written as:
\begin{math}
\label{bcs}
A_{BCS}(k,\omega)=\frac{1}{\pi}[\frac{|u_{k}|^2\Gamma}{(\omega-E_{k})^2+\Gamma^2}+\frac{|v_{k}|^2\Gamma}{(\omega+E_{k})^2+\Gamma^2}],
\end{math}
where u$_{k}$ and v$_{k}$ are coherence factors, $\Gamma$ is a linewidth parameter, and E$_{k}$=($\epsilon_k^2+|\Delta(k)|^{2})^{1/2}$ where $\epsilon$(k) is the normal state band dispersion and $\Delta$(k) is the gap function. Two prominent characteristics are noted in the simulated spectral function in the superconducting state (Fig. la). The first is the existence of two branches of Bogliubov quasiparticle dispersions above and below the Fermi level E$_F$, separated by the superconducting gap 2$|\Delta|$ and satisfying a relation A(k,$\omega$)=A(-k,-$\omega$).  The second is that, for a given branch, there is a band back-bending right at the Fermi momentum k$_F$.  Since ARPES measures a single particle spectral function A(k,$\omega$) weighted by the photoemission matrix element and the Fermi distribution function f($\omega$,T):
\begin{math}
I(\mathbf{k},\omega)=I_0(\mathbf{k},\nu,A)f(\omega,T)A(\mathbf{k},\omega),
\end{math}
it probes mainly the occupied states due to $f(\omega,T)$. When the temperature (T) is sufficiently low, the upper branch of the dispersion is suppressed (Fig. 1c).  On the other hand, when the temperature is relatively high in the superconducting state, a few electronic states  above the Fermi level can be thermally populated and become visible (Fig. 1b).

\begin{figure}[b]
\begin{center}
\includegraphics[width=1.00\columnwidth,angle=0]{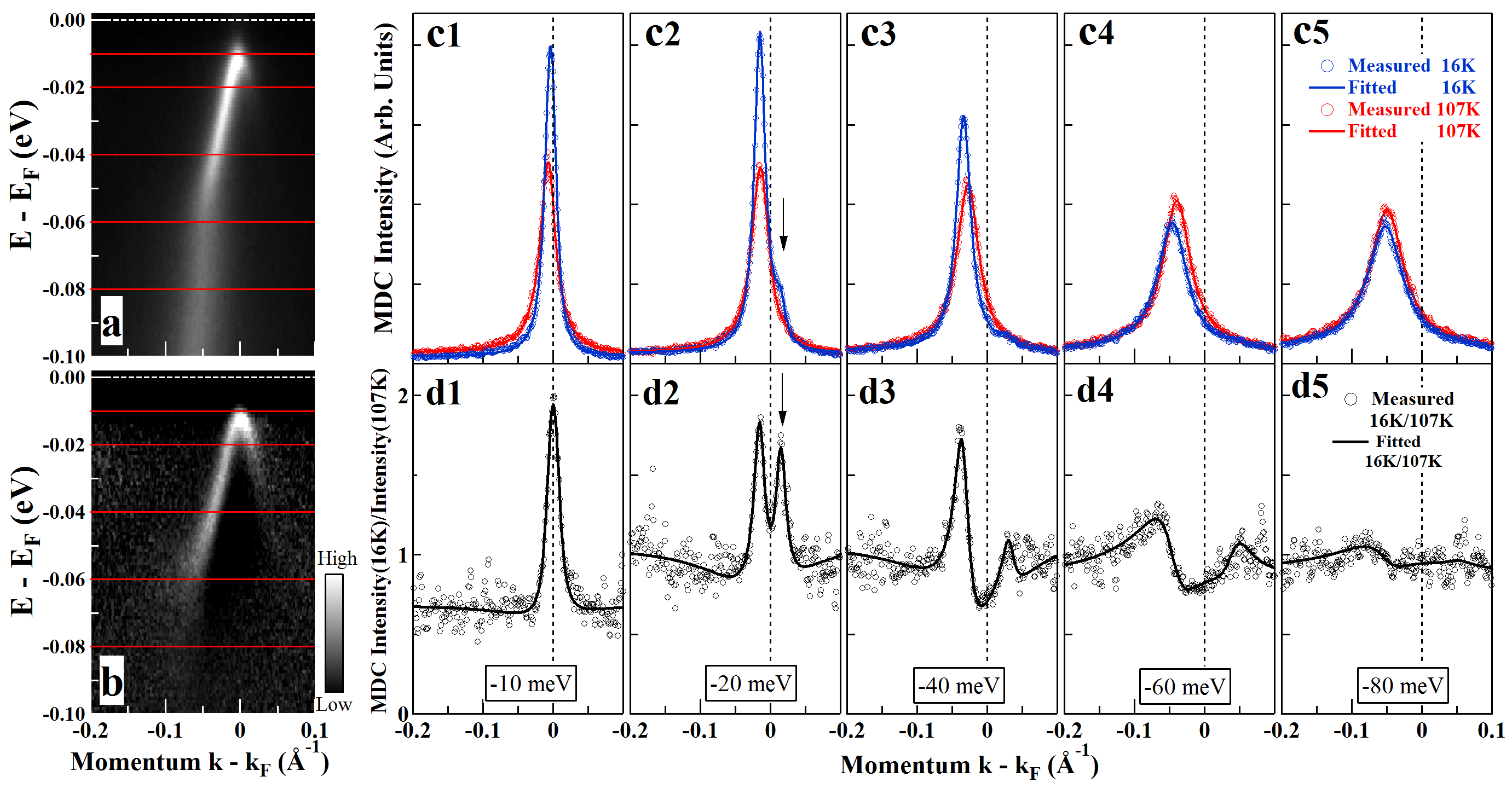}
\end{center}
\caption{Observation of band back-bending in the lower branch of the Bogoliubov quasiparticle dispersion in the superconducting state of Bi2212. (a). Photoemission image taken at 16 K along the cut shown in Fig. 1d. (b). Photoemission image taken at 16 K divided by the image taken at 107 K along the same cut. (c1-c5). Representative MDCs in the normal state at 107 K (red circles) and superconducting state at 16 K (blue circles) at five different binding energies. The solid lines represent fitted results.   (d1-d5). MDCs at 16 K divided by MDCs at 107 K at different binding energies (black circles).
}
\end{figure}

The first characteristic of the Bogoliubov quasiparticle, i.e., the existence of two dispersion branches above and below the Fermi level in the superconducting state, has been reported in the previous ARPES measurements on cuprate superconductors\cite{HMatsui_Bi2223,AVB_BogoAngle,WSLee,HBYang_CooperPair}. It shows up more clearly in our super-high resolution ARPES measurements (Fig. 1(e1-e5)).  As seen in Fig. 1e1, at low temperature (16 K), the spectral weight above the Fermi level is almost invisible. However, at relatively high temperatures while the sample remains in the superconducting state, there are clear features present above the Fermi level, as seen from  the 70 K (Fig. 1e2) and 80 K (Fig. 1e3) measurements.  By dividing out the corresponding Fermi distribution function from the measured photoemission data in Fig. 1(e1-e5), part of the upper dispersion branch near the Fermi level can be recovered, as seen in the 70 K (Fig. 1f2) and 80 K (Fig. 1f3) data.  Indeed, the upper branch and the lower branch are nearly centro-symmetric with respect to the Fermi momentum at the Fermi level.   Above T$_c$, the band recovers to the normal state dispersion with no gap opening (97 K data in Figs. 1e4 and 1f4, and 107 K data in Figs. 1e5 and 1f5).

\begin{figure}[t]
\begin{center}
\includegraphics[width=1.00\columnwidth,angle=0]{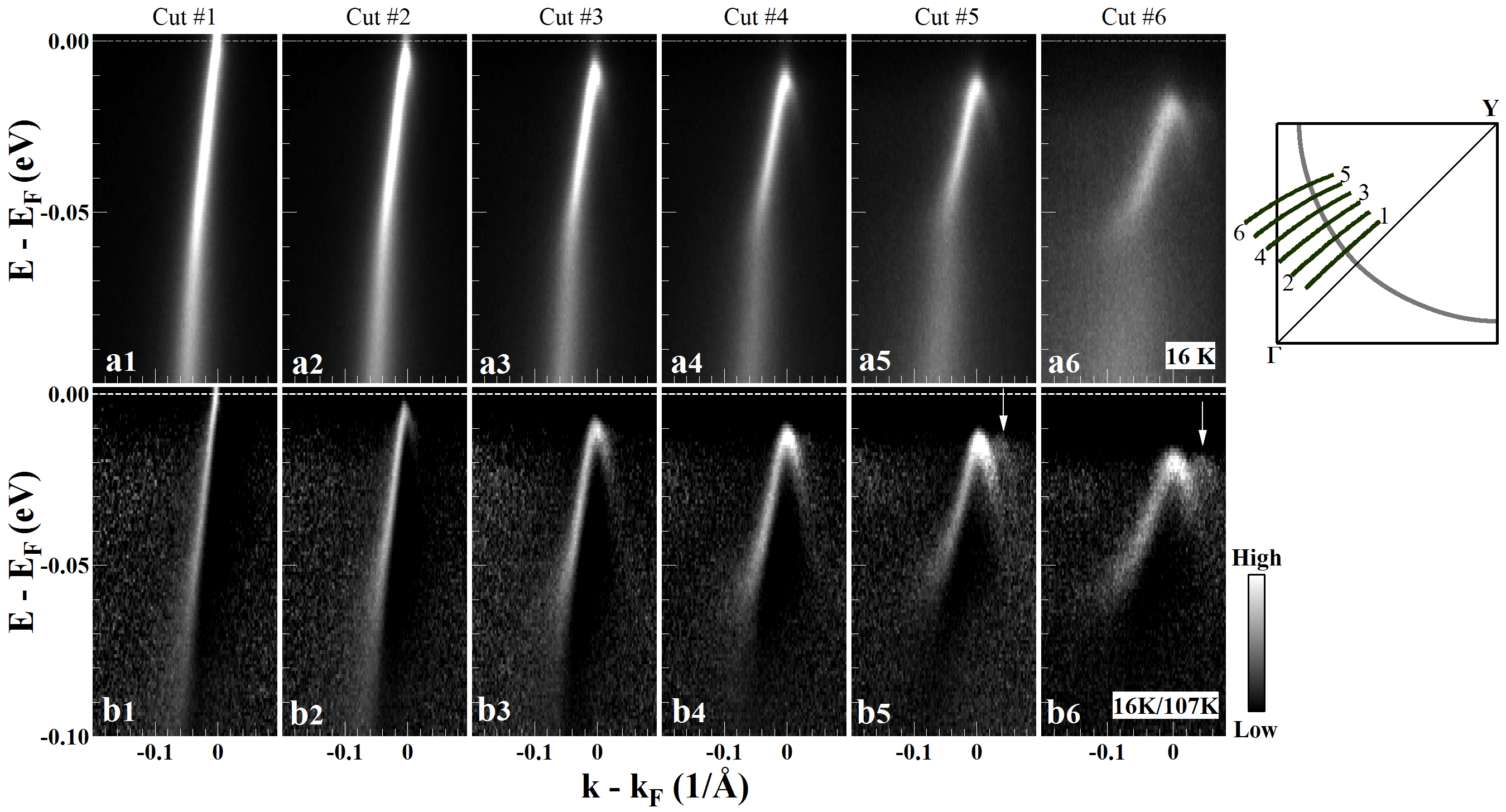}
\end{center}
\caption{Momentum dependence of the lower branch of the Bogoliubov quasiparticle-like dispersion in Bi2212. (a1-a6). Photoemission images measured at 16 K for different momentum cuts with their location shown in the upper-right inset. (b1-b6). Photoemission images measured at 16 K divided by the corresponding  images measured at 107 K. The white arrows in (b5) and (b6) mark an additional feature that is likely due to umklapp bands in Bi2212.
}
\end{figure}

One major result of the present work is the revelation of the other characteristic of the Bogoliubov quasiparticles in the superconducting state, i.e.,  the band back-bending expected in the lower branch at the Fermi momentum (Fig. 1c). In this case, a momentum distribution curve (MDC) (red thick line in Fig. 1c) at an energy slightly below the gap $\Delta$ (red thin line in Fig. 1c) is expected to exhibit two peaks: a strong peak from the original main band, and the other weak peak (or shoulder, marked by black arrow in Fig. 1c) from the superconductivity-induced additional band. Indeed, as seen from Fig. 2a which was taken on Bi2212 at 16 K in the superconducting state, a band back-bending behavior can be clearly observed.  The MDC at a typical binding energy of 20 meV shows a clear shoulder in addition to the main peak (Fig. 2c2).  To highlight the superconductivity-induced change across the superconducting transition, we also show in Fig. 2b the photoemission image taken at 16 K divided by the one taken at 107 K in the normal state. The superconductivity-induced band back-bending becomes more pronounced (Fig. 2b) and the corresponding MDCs at some binding energies like 20 meV show two clear peaks (Fig. 2d2). To the best of our knowledge, this is the first time that this second characteristic of the Bogoliubov quasiparticles in the superconducting state is revealed so clearly, mainly due to the much improved resolution of our laser-based ARPES measurements.

\begin{figure}[b]
\begin{center}
\includegraphics[width=1.00\columnwidth,angle=0]{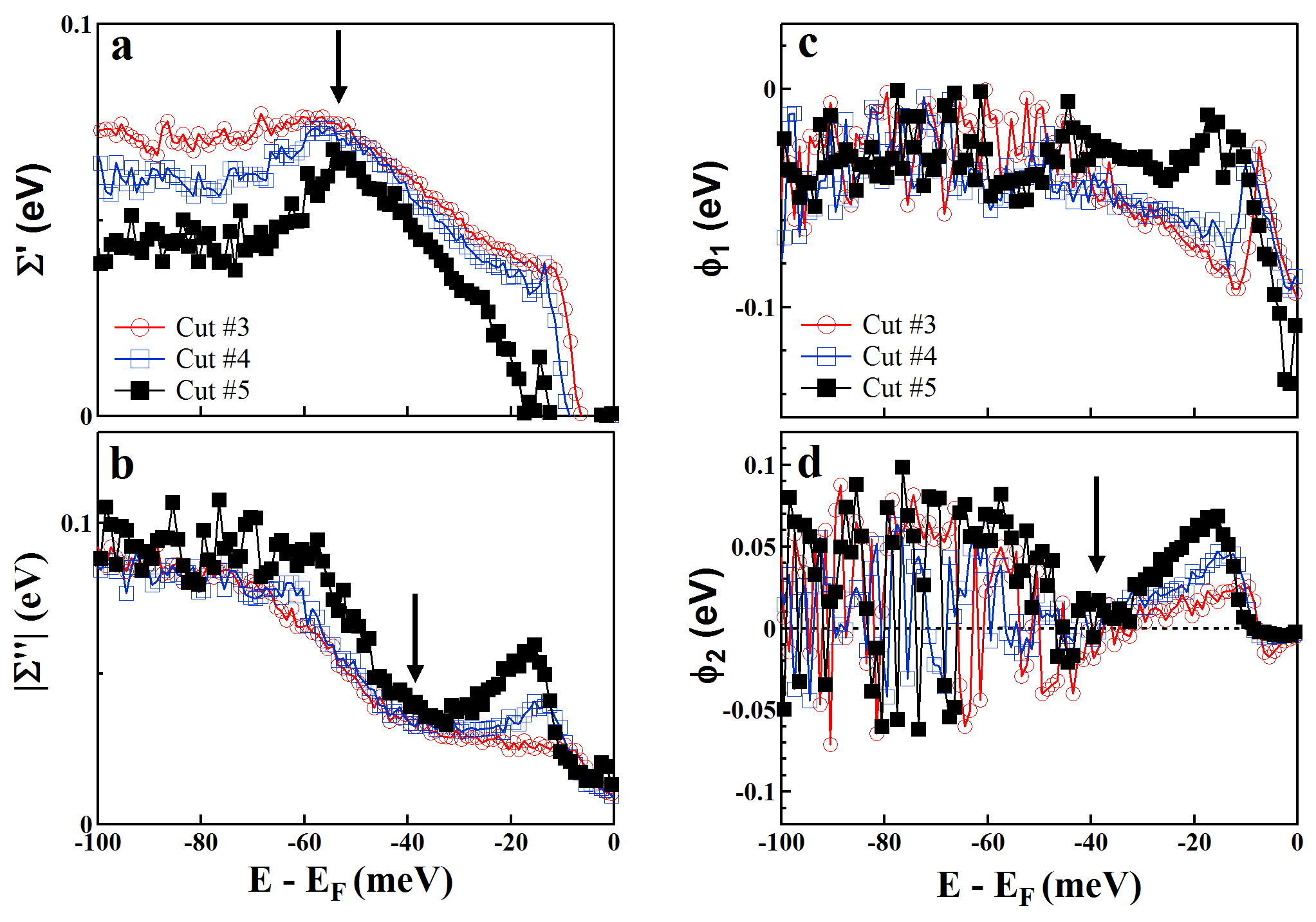}
\end{center}
\caption{The electron self-energy and gap function obtained from the ARPES measurements of Bi2212 in the superconducting state. (a) and (b) show real part and imaginary part, respectively, of the electron self-energy $\Sigma(\omega)$ obtained for three different momentum cuts (Cuts $\#$3, $\#$4 and $\#$5 in upper-right inset of Fig. 3); (c) and (d) show the real part and imaginary part, respectively, of the gap function $\phi(\omega)$ for the three momentum cuts.}
\end{figure}

Fig. 3(a1-a6) show the momentum dependence of the photoemission images measured at 16 K;  the corresponding photoemission images at 16 K divided by those at 107 K are shown in Figs. 3(b1-b6) to highlight the change across the superconducting transition.  The band back-bending behavior is clearly observed and appears to get more pronounced as the momentum cuts move away from the nodal region towards the antinodal region.  The energy position of the bending points moves to a larger binding energy as the momentum cuts move towards the antinodal region; this is consistent with the increasing superconducting gap size because the location of the band-bending top corresponds to the superconducting gap $\Delta$ at the Fermi momentum (Fig. 1a).

The observation of the two major characteristics of the Bogoliubov quasiparticle-like behavior in Bi2212 indicates that its superconducting state is qualitatively consistent with the BCS formalism. More importantly, the clear revelation of the band back-bending  behavior at low temperature offers a new opportunity to go beyond the BCS approximation by using the Eliashberg formalism to directly extract the electron self-energy and the gap function from the ARPES data. The single particle spectral function A(k,$\omega$), measured by ARPES, is proportional to the imaginary part of the Green's function
\begin{equation}
A(\mathbf{k},\omega)=-\frac{1}{\pi}Im\{G(\mathbf{k},\omega)\}
\label{Akw}
\end{equation}
Considering a cylindrical Fermi surface in the superconducting state, it can be written as
\begin{equation}
G(k,\omega)=\frac{Z(\omega)\omega+\epsilon_k}{(Z(\omega)\omega)^2-\epsilon^2_k-\phi^2(\theta,\omega)}
\label{GreenF}
\end{equation}
where $\epsilon_k$ is the bare band, $Z(\omega)$ is a renormalization parameter, and the $\phi(\theta,\omega)=\phi(\omega)\cos(2\theta)$ is a {\it d}-wave gap function. This form is applicable when the momentum cut points towards  $(\pi, \pi)$, as is the case for our measurements (see Fig. 1d and inset of Fig. 3).  Z$(\omega)$ and $\phi(\omega)$ are extracted by fitting the MDCs at different binding energies using Eqs.~(\ref{Akw}) as shown in Fig. 2c.  In the fitting procedure, the bare bands $\epsilon_k$ are taken from the tight binding model\cite{RSM_TightBinding} which are the same as used before\cite{HYPRB}. Different selection of the bare bands would affect the absolute value of the fitted quantities (Fig. 4) but has little effect on the main features that will be discussed below.  All the MDCs are well fitted by the combined Eqs. (\ref{Akw}) and (\ref{GreenF}), as shown in Fig. 2c for MDCs (16 K, blue lines and circles) at several typical binding energies.  The same equations also fit the normal state data by taking $\phi(\omega)$=0 in Eqs.~(\ref{GreenF}), as shown in Fig. 2c for the 107 K data\cite{HYPRB}.

Fig. 4 shows the obtained real and imaginary parts, $\Sigma^{\prime}(\omega)$ and $\Sigma^{\prime\prime}(\omega)$, of the electron self-energy,  and the real and imaginary parts, $\phi_1(\omega)$ and  $\phi_2(\omega)$, of the gap function for three typical momentum cuts (cuts $\#$3,  $\#$4, and  $\#$5 in upper-right inset of Fig. 3).  While the gap function $\phi(\omega)$ is obtained directly from the above fitting procedure, the electron self-energy $\Sigma(\omega)$ is obtained from the fitted renormalization parameter Z($\omega$) by $\Sigma(\omega)=[1-Z(\omega)]\omega$. Since the superconductivity-induced change occurs most obviously in a small energy range near the Fermi level (Fig. 3b), we confine our fitting results within 100 meV energy window near the Fermi level.  The features below $\sim$20 meV are mainly related to the opening of superconducting gap (for these three cuts, the corresponding superconducting gap is between 10 and 15 meV).  At higher energies, two main features can be identified: one at $\sim$54 meV showing as a robust hump in the real part of the electron self-energy (Fig. 4a), and the other at $\sim$40 meV showing as  a dip in both the imaginary part of the electron self-energy (Fig. 4b) and the imaginary part of the gap function (Fig. 4d).  We note that the $\sim$54 meV feature is close to the bosonic mode observed in the tunneling experiment\cite{DavisSTM} and is also close to the energy scale of the well-known nodal dispersion kink in cuprates\cite{NodalKink}. The 40 meV feature is close to the antinodal kink found in Bi2212, with its energy close to either the resonance mode or the B$_{1g}$ phonon mode\cite{TCuk}.  Further work are needed to pin down the exact origin of these energy scales and their role in causing superconductivity. We also note that $\Sigma(\omega)$ at higher energies (above 50 meV) show a featureless background that is also observed in the normal state\cite{HYPRB}.

In summary, by taking advantage of the high precision ARPES measurements on Bi2212, we have resolved clearly both characteristics of the Bogoliubov quasiparticle-like dispersions in the superconducting state. In particular, the revelation of the band back-bending behavior of the lower dispersion branch  at low temperature makes it possible for the first time to extract the complex electron self-energy and complex gap function of Bi2212 superconductor in the superconducting state. The experimental extraction of the electron self-energy and the gap function in the superconducting state will provide key information and constraints on the pairing mechanism in high temperature superconductors. First, like in the conventional superconductors, it can provide examinations  on various pairing theories by computing these two quantities to compare with the experimentally determined ones. Second, also like in the conventional superconductors, if it is possible to directly perform inversion of these two quantities to obtain the underlying bosonic spectral function that is responsible for superconductivity, it may provide key information  on the nature of the electron pairing mechanism. We hope our present work will stimulate further efforts along these directions.

XJZ thanks the funding support from NSFC (Grant No. 10734120) and the MOST of China (973 program No: 2011CB921703).

$^{*}$Corresponding author (XJZhou@aphy.iphy.ac.cn)

\end{document}